\documentclass[10pt,tightenlines,eqsecnum,floats,aps,amsmath,amssymb,nofootinbib,superscriptaddress,prd,showpacs,twocolumn]{revtex4}

\usepackage{graphicx}

\usepackage{euscript,amssymb}

\usepackage{epsfig}
\usepackage{color}

\usepackage{enumitem}

\usepackage{amsfonts}

\usepackage{amsmath}

\usepackage{amssymb}

\usepackage{fancyhdr}

\usepackage{esint}
\usepackage[unicode=true, pdfusetitle,
 bookmarks=true,bookmarksnumbered=false,bookmarksopen=false,
 breaklinks=false,pdfborder={0 0 1},backref=false,colorlinks=false]
 {hyperref}

\begin{document}

\title{Creation of particles in a cyclic universe driven by loop quantum cosmology}

\author{Yaser Tavakoli}
\email{tavakoli@cosmo-ufes.org}
\affiliation{Departamento de F\'isica, Universidade Federal do Esp\'irito Santo, Av. Fernando Ferrari 514, 29075-910 Vit\'oria - ES, Brazil}

\author{J\'ulio C. Fabris}
\email{fabris@pq.cnpq.br}
\affiliation{Departamento de F\'isica, Universidade Federal do Esp\'irito Santo, Av. Fernando Ferrari 514, 29075-910 Vit\'oria - ES, Brazil}

\begin{abstract}

We consider an isotropic and homogeneous  universe in loop quantum cosmology.  
We assume that the  matter content of the universe is  dominated by dust matter  in early time and  
a phantom matter at late time which constitutes  the dark energy component.
The quantum gravity  modifications to the Friedmann equation  in this  model
indicate that  the classical  big bang singularity and the future  big rip singularity 
are resolved  and are replaced  by quantum bounce.
It turns out that  the big bounce and recollapse in the herein model  contribute to  a cyclic scenario for the universe. 
We then  study the  quantum  theory  of a massive, non-minimally coupled scalar field  undergoing cosmological evolution 
from  primordial  bounce  towards the late time  bounce.
In particular, we solve the Klein-Gordon equation for  the  scalar field in  the primordial and late time regions,
in order to investigate particle production phenomena  at late time.  
By  computing    the  energy density of created particles at late time,
we  show  that  this  density  is   negligible in comparison to   the  quantum  background density at
Planck era. This indicates that  the effects of  quantum particle production do  not influence  the  future bounce.

\end{abstract}

\date{\today}


%

\maketitle


\section{Introduction}
\label{intro}

Quantum field theory (QFT) in curved space-time is the theory of quantum
fields propagating on a classical background \cite{Birrell, mw, Fulling:1989}. 
This theory has provided a good approximate description of quantum phenomena  in a regime where the
quantum effects of gravity do not play a dominant role, but  the effects of curved space-time may
be significant. In particular, this theory was applied to the description of quantum phenomena 
occurring in the early universe or close to the black holes.
Nevertheless, in the regimes arbitrarily close to the classical singularities 
where the space-time curvatures reach Planckian scales,
quantum effects of gravity cannot be
neglected, hence, the theory of quantum fields in classical curved space-time is no longer valid.
Therefore, it is expected that  
the quantum nature of the space-time would have to be taken into account 
while studying the QFT in Planck era.

The  theory of quantum gravity is one of the major open problems in physics, 
even though by now there are some interests in loop quantum gravity (LQG) \cite{alrev,ttbook,crbook}. 
Loop quantum cosmology (LQC),  is one possible approach to investigating quantum gravity effects
 in the early universe  by using the quantization techniques from LQG,
which provides a number of concrete results \cite{abl,aa-badhonef,Bojowald:livingRev,Bojowald:LQC,Ashtekar-Singh}.
In particular, LQC predicts that  the quantum  modification of space-time geometry
at early times replaces the big bang singularity by a big bounce when the energy density of the universe
approaches the Planck energy \cite{aps1,aps2,aps3}.
While the model shows the  key role played by quantum effects in resolution of big bang singularity at early time,
the question whether quantum gravity effects  can also be manifested in large scale cosmology has been 
investigated  in Refs. \cite{Sami:2006,Singh:2011,Ding:2011}.

According to several astrophysical observations, our universe is  undergoing a state of accelerating
expansion.  Such experiments indicate that the matter content of the universe, leading to the accelerating expansion,
must contain dark energy component 
(with equation of state 
$p_{\rm DE}=w_d\rho_{\rm DE}$ where $w_d<-1/3$) which  constitutes $68\%$ of the total matter content of the universe \cite{Planck}. 
The recent astrophysical data as observed by the {\em Planck} (using baryon acoustic oscillation (BAO) and  Wilkinson Microwave Anisotropy Probe
polarization low-multipole likelihood (WP) data), for a constant $w_d$ and a universe with the  flat spatial section,
indicates that 
$w_d = -1.13^{+0.24}_{-0.25}$  at $2\sigma$  \cite{Planck}.  
From now on, we will thus suppose  that dark energy is {\em phantom}, 
that is $w_d<-1$.
When the universe is expanding, the density of dark
matter decreases more quickly than the density of dark energy, 
and eventually the matter content of the universe becomes dark energy dominant at late times.
Therefore, dark energy might  play an  important  role on the  implications for the fate of the universe \cite{singularities-latetime}.
In the context of dark energy cosmology, there have been many classical  investigations into the
possibility that expanding universe  can come to a violent end at a finite cosmic time, experiencing a singular fate.  However, 
the effects of LQC  correction might  impose an upper bound for  the  density of dark energy at late time,
that  lead  to resolving the future singularities and replacing them by a quantum bounce.
Interestingly, this indicates that the scale factor of the universe undergoes contracting and expanding phases  periodically,
so that the universe  can possess an exactly cyclic evolution \cite{Ding:2011,Sami:2006}.

One restriction in experiencing the quantum gravity is that the Planck energy scale is very far  
beyond the reach of standard experiments such as particle accelerators. 
Ultra-high energy phenomenas,  possibly capable of probing the Planck scale,  
have occurred  at early universe. 
Although  we cannot  re-do the early universe, we can witness its consequences.
While the theory of quantum fields propagating on classical FLRW background is well-known, 
there has been some attempts  to develop the theory of test quantum fields propagating on a quantum cosmological space-time 
where the  background geometry is  governed  by the LQC model
of the homogeneous and isotropic universe \cite{AKL,DLT:2012,Ashtekar:2013,Andrea:2013}. 
These studies have also provided an extension of the quantum theory of cosmological perturbations to the Planck era \cite{Ashtekar:2013}.
Furthermore, by a similar study of  the  QFT on a Bianchi I quantum space-time, 
it  has been provided  some phenomenological  insight  into the effects of the quantum nature of geometry 
on the propagation of test fields and the possible violation of the local Lorentz symmetry \cite{DLT:2012}.
These effects might be possibly  observed  in some cosmological experiments.
A general prediction of the QFT in a curved background at early universe is that particles can be created by time-dependent  gravitational fields \cite{Parker:1966,Parker:1969,Parker:1974,Pauli-Villars,Grib:1994,Grib:1998}.
Interestingly,  a cyclic universe in LQC which initiated  from a big bounce and then expanding  towards a future bounce, 
may  provide a  background with no unique vacuum state. This  implies  a  mechanism  for quantum particle production,
and  further  provides a circumstance  in which one can  measure and observe interesting 
QFT phenomena at Planck era   in an expanding universe.
This constitutes our main goal to investigate  within  this paper.

In this paper we study the quantum fields propagating on a cosmological  background  governed by the  improved dynamics framework LQC.
The paper is organized as follows. 
In section \ref{lqc} we consider a Friedmann-Lema\^itre-Robertson-Walker (FLRW) universe dominated by  a dust matter in the past and a phantom dark energy component in the future; 
then, by employing LQC modifications to the Friedmann equation, we show that,  the classical big bang and  big rip singularities are resolved 
 and are replaced by quantum bounces at Planck era, one in early time and the other at late time. 
It will be further shown that such expanding and recollapsing phases lead  to a cyclic behaviour  for the universe.
In section \ref{QFT} we study the QFT  in
quantum cosmological background   we presented in section \ref{lqc}.
We will then pay a particular attention on the mechanism of particle production
in the universe at late time. 
Finally,  we will present the conclusions and  the results of our work in section \ref{conclusion}.

\section{Quantum cosmological scenario}
\label{lqc}

We consider a FLRW universe whose matter content constitutes a dust  matter with the  density $\rho_{\rm matt}$, 
and a dark energy with the density $\rho_{\rm DE}$;
its  total energy  density $\rho$ reads
\begin{align}
\rho\ =\ \rho_{\rm matt} + \rho_{\rm DE}\ .
\end{align}
The total energy density  $\rho$  must satisfy the conservation equation:
\begin{align}
\dot{\rho}+3H(\rho+p)=0\ .
\label{conservation}
\end{align}
Furthermore, we assume that there is no interaction between dark matter and dark energy components, 
so that, each component must also satisfy the local conservation equation:
\begin{align}
& \dot{\rho}_{\rm matt}+3H\rho_{\rm matt}\ =\ 0 \ , \label{loc-energy1} \\
& \dot{\rho}_{\rm DE}+3H(\rho_{\rm DE}+p_{\rm DE})\ =\ 0 \ .
\label{loc-energy2}
\end{align}

In effective loop quantum cosmological scenario, evolution equations of the universe are given by the modified Friedmann equation \cite{aps3}
\begin{align}
H^2\ =\ \frac{\kappa}{3}\rho\left(1-\frac{\rho}{\rho_{\rm crit}}\right)\ ,
\label{Friedmann}
\end{align}
and the Raychaudhuri equation
\begin{align}
\dot{H}\ =\  -\frac{\kappa}{2}(\rho+p)\left(1-2\frac{\rho}{\rho_{\rm crit}}\right)\ ,
\label{timederive-Friedmann}
\end{align}
where $\kappa:=8\pi G$ and $\rho_{\rm crit}=0.41\rho_{\rm Pl}$ is the upper 
bound for the energy density of the universe provided by quantum gravity  ($\rho_{\rm Pl}$ is the Planck energy density) \cite{aps3}.

Our  goal in  this section is  to construct  a cyclic cosmological scenario in  LQC  (see for example \cite{Sami:2006}). 
Therefore, we consider an expanding phase for the universe initiated at 
a preferred  instant of time, $t_i=0$, being the initial
quantum bounce  resulted from loop-quantum-gravity effects  at early time \cite{aps1}. 
Then, in the far future when the  density of the dark energy components  becomes large and  comparable with the Planck energy,
modifications to the Friedmann equation given in LQC, again give rise to a quantum bounce for the fate of the universe. 
Within this scenario, we can select  two cosmological phases which, as in QFT, 
can  play the role  of  `in' and `out'  regions  for the quantum (scalar) fields propagating on the 
quantum  background;  they are, respectively,  the `initial' and `late time' bounces. 

In this model we assume that  the matter content of the universe  is dominated by dust matter ($\rho\approx\rho_{\rm matt}$) at early time,
and is dominated by  dark energy with phantom component (with the density $\rho\approx\rho_{\rm DE}$) at late time.  
We consider that  transition from matter dominant  phase  to dark energy dominant  phase occurs  at  the present time $t=t_0$.
Let us describe these two cosmological regimes in more details:
\begin{enumerate}[label=(\roman*)]  
\item  In the region  $0<t<t_0$,  
we assume that  the matter content of the universe  is dominated by
 a dust fluid whose  density is $\rho_{\rm matt}$ with the pressure $p_{\rm matt}=w_M\rho_{\rm matt}=0$. Then, from Eq.~(\ref{loc-energy1})
 we obtain 
\begin{align}
\rho (a)\ \approx\ \rho_{\rm matt}\ =\ \rho_{\rm crit}\left(\frac{a}{a_i}\right)^{-3}\ .
\label{energy-primordial}
\end{align} 
In this region,  we assume that at some very early time $t=t_i=0$,
the universe begins to expand; in other words, at $t_i=0$, we have that $\dot{a}(t=0)=0$ and $\ddot{a}(t=0)>0$.
Therefore, from the modified Friedmann equation (\ref{Friedmann}) we have that 
the energy density of the universe has its maximum value $\rho(0)=\rho_{\rm crit}$ at $t=0$,
where it is at a minimum scale $a(0)=a_i$.
Then, for $t>0$ the energy density of the universe (i.e., dust component) decreases until present time $t=0$,
when the energy density and the scale factor of the universe are  $\rho(t_0)=\rho_0$ and $a(t_0)=a_0$, respectively.
By integrating the Friedmann equation (\ref{Friedmann}), we obtain 
time evolution of the energy density  in this region as 
\begin{align}
\rho(t) \  =\  \rho_{\rm crit}\left(1+\frac{3}{4}\kappa\rho_{\rm crit} t^2\right)^{-1}.
\label{energy-primordial-B}
\end{align}
Then, by substituting this in Eq.~(\ref{energy-primordial}),
 time evolution of the scale factor $a(t)$ of the universe in this regime is obtained as: 
\begin{align}
a(t) \  =\  a_i\left(1+\frac{3}{4}\kappa \rho_{\rm crit} t^2\right)^{1/3}.
\label{scalefactor1}
\end{align}
\item In the region   $t_0<t<t_b$,  the universe becomes  dark energy dominant:  $\bar{\rho}:=\rho(t>t_0)\approx \rho_{\rm DE}$.
The dark energy component is  assumed to be a fluid satisfying the
equation of state  $p_{\rm DE}=w_{d}\rho_{\rm DE}$,  where $w_{d}$ is a constant.
Then, from  equation of conservation (\ref{loc-energy2})
we find  the  density of dark energy component as
\begin{align}
\bar{\rho}\ \approx\ \rho_{\rm DE}\ =\ \rho_{0}\left(\frac{\bar{a}}{a_0}\right)^{-\beta}\ ,
\label{energy-latetime}
\end{align}
where $\bar{a}(t):=a(t>t_0)$ and $\beta=3(1+w_d)<0$ (for a phantom fluid with $w_d\lesssim-1$). 
This implies that  the  energy density of the universe, in this region, increases until it reaches an upper bound $\rho_{\rm crit}$ in the future.
By integrating the Friedmann equation (\ref{Friedmann}) for this case, we find the time dependent energy density (\ref{energy-latetime}) as
\begin{align}
\quad \quad  \bar{\rho}(t) \  =\  \rho_{\rm crit}\left[1+\frac{\kappa}{12}\rho_{\rm crit}\beta^2(t_b-t)^2\right]^{-1}.
\label{energy-latetime-B}
\end{align}
This equation indicates that,
in the far future, at some times $t=t_b$  
the energy density of the universe reaches its maximum $\bar{\rho}(t_b)=\rho_{\rm crit}$ and the universe bounces;  $\dot{a}(t_b)=0$. 
By replacing $\bar{a}(t_b)=a_b$, at which $\bar{\rho}=\rho_{\rm crit}$, in Eq.~(\ref{energy-latetime}), we obtain the scale factor $a_b$ of the universe at the future bounce:
\begin{align}
a_b\ =\ a_0\left(\frac{\rho_{\rm crit}}{\rho_0}\right)^{\frac{1}{|\beta|}}.
\label{scalefactor2}
\end{align}
Consequently, we can obtain the evolution of the scale factor  in the late time era by substituting $\bar{\rho}(t)$  in  (\ref{energy-latetime}) from  Eq.~(\ref{energy-latetime-B}):
\begin{align}
\quad \quad \bar{a}(t) \  =\  a_{b}\left[1+\frac{\kappa}{12}\rho_{\rm crit}\beta^2(t_b-t)^2\right]^{-\frac{1}{|\beta|}}.
\label{scalefactor2-b}
\end{align}
By setting $\bar{\rho}(t_0)=\rho_0$ in Eq.~(\ref{energy-latetime-B}) at present time,  the time $t_b$ at which the universe hits a bounce in the future is determined:
\begin{align}
t_b \  &=\  t_0 + \sqrt{\frac{12}{\kappa\rho_{0}\beta^2}\left(1-\frac{\rho_0}{\rho_{\rm crit}}\right)} \notag \\
& =\ t_0 + \frac{6H_0}{\kappa\rho_{0}|\beta|}\ .
\label{t-b}
\end{align}
Considering  that at present time $t=t_0$ we have $\rho_0/\rho_{\rm crit}\ll 1$, the Friedmann equation  (\ref{Friedmann})  can be approximated as  $H_0^2\approx\kappa\rho_0/3$. Putting this in Eq.~(\ref{t-b}) we obtain
\begin{align}
t_b  \ \approx\   t_0 + \frac{2}{|\beta| H_0} \  .
\label{t-b2}
\end{align}
In this equation, by setting $t_0$  as  $t_0\sim1/H_0$, we obtain  $t_b  \sim (1+2/|\beta|)t_0$, indicating that
the universe will hit a bounce in some time $(2/|\beta|)t_0$ in the future.
\end{enumerate}
From the matching condition at $t=t_0$ for the two (past  and future) cosmological regions  
we have that the time derivative of the scale factors  satisfy   $\dot{a}(t_0)=\dot{\bar{a}}(t_0)$ which corresponds to 
$H(t_0)=\bar{H}(t_0)=H_0$.
Suppose  that  $\rho_0/\rho_{\rm crit}\ll 1$ at present time,
 the Hubble rate takes its classical limit,  $H_0^2\approx\kappa\rho_0/3$.
 This leads to the matching of  energy components of the universe in two cosmological regions: $\rho(t_0)=\bar{\rho}(t_0)=\rho_0$.

\begin{figure}[t]
\includegraphics[height=2.1in]{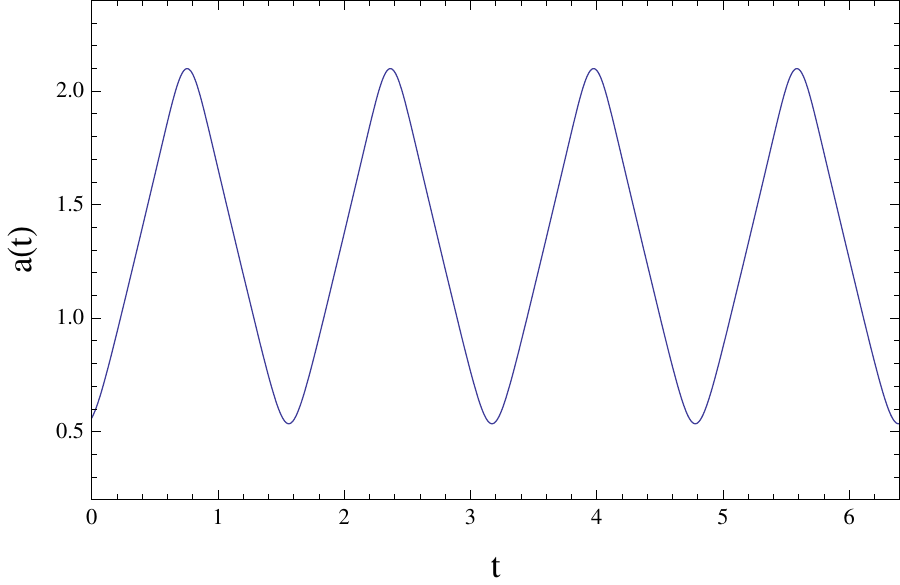} 
\caption{Numerical solution  for  the scale factor of the universe, governed  by  Eqs. (\ref{Friedmann}) and (\ref{timederive-Friedmann}), 
for the rescaled  choices of parameters $\beta=3(1+w_d)=-6.6$, $\rho_0/\rho_{\rm crit} = 0.005$, $a_i=0.56$ and $\dot{a}(0)=1$. 
This shows an oscillatory behaviour for the scale factor in the whole evolution of the universe.
}
\label{fig-scale} 
\end{figure}

Figure \ref{fig-scale} presents  a  numerical    solution for   the scale factor  $a(t)$ of the universe in our model,
considering simultaneously both fluids, 
where  $a(t)$  oscillates  in the region $a_i<a<a_b$ between the primordial and late time bounces.
Notice that the values we have used  for the parameters  are not necessarily realistic; they were chosen for a better
visualisation of the scenario.

\section{Quantum field theory}
\label{QFT}

We consider the cyclic cosmological background, resulted from LQC,  which was presented in the previous section. 
In this section, we study the propagation of quantum fields on this background space-time.
Then, we investigate circumstances for creation of quantum particles near the future  bounce.

\subsection{Scalar field  on the quantum corrected FLRW background space-time}

Let us consider a real (inhomogeneous)  scalar field $\phi(t,\vec{x})$ on a   FLRW  background, whose Lagrangian is
\begin{eqnarray}
\mathcal{L}_{\phi}\ =\  -\frac{1}{2} \left(g^{ab} \partial_{a} \phi \partial_{b} \phi + m^{2} \phi^{2} + \xi {\cal R}\phi^2\right),
\label{26}
\end{eqnarray}
where ${\cal R}$ is the curvature of the gravitational system; the coupling  constants  $\xi=0$ and $\xi=1/6$ denote respectively, to  the cases of {\em minimally} and {\em non-minimally} coupled scalar fields.
By variation of the Lagrangian above with respect to $\phi$, we obtain the Klein-Gordon equation
\begin{align}
\left(\square - m^2 - \xi{\cal R}\right)\phi(t, \vec{x})\ =\ 0\ .
\label{KG1}
\end{align}
Performing the Legendre transformation, one gets the canonically conjugate momentum for the test field $\phi$, denoted by $\pi_\phi$, on a $t = const$ slice.
Then, for the pair $(\phi,\pi_\phi)$, the classical solutions of the equation of motion (coming from (\ref{26})) can be expanded in Fourier modes:
\begin{align}
 \phi(t, \vec{x})\  & =\  \frac{1}{(2\pi)^{\frac{3}{2}}} \int d^3k~ \phi_{\vec{k}}(t) e^{i \vec{k} \cdot \vec{x}}, \label{27-a} \\
\pi_{\phi}(t, \vec{x})\ &  =\  \frac{1}{(2\pi)^{\frac{3}{2}}} \int d^3k~ \pi_{\vec{k}}(t) e^{i \vec{k} \cdot \vec{x}},
\label{27}
\end{align}
where the Fourier coefficients $\phi_{\vec{k}}(t)$ and $\pi_{\vec{k}}(t)$ satisfy the commutation relation
$\{\phi_{\vec{k}}, \pi_{\vec{k}'}\} = \delta (\vec{k} + \vec{k}')$, and the reality conditions
$\phi_{\vec{k}}^\ast = \phi_{-\vec{k}}$ and $\pi_{\vec{k}}^\ast = \pi_{-\vec{k}}$ \cite{mw}.

In the conformal form, the FLRW metric reads
\begin{eqnarray}
g_{ab} dx^{a} dx^{b}\ =\  C(\eta)\left(-d\eta^{2} +  d\vec{x}^{2}\right)\ ,
\label{metric-conformal}
\end{eqnarray}
with $\eta$ being the {\em conformal time} where $C^{1/2}(\eta)d\eta=dt$.
For convenience, we  introduce the auxiliary field given by $\chi := a \phi$, 
so that,  each mode of the form  $\phi_{\vec{k}} := \chi_{\vec{k}}/a$ satisfies the equation
\begin{align}
\chi_{\vec{k}}'' + \Omega_{k}^2(\eta)\chi_{\vec{k}}=0\ .
\label{Master-Eq1}
\end{align}
A `prime' denotes to the  derivative with respect to the conformal time $\eta$. 
The frequency $\Omega_{k}$ is given by \cite{Birrell}
\begin{align}
\Omega_{k}^2(\eta)\ :=\ k^2+C(\eta)\left[m^2 + (\xi -1/6) {\cal R}\right]\ ,
\label{Frequency}
\end{align}
where $k^2=|\vec{k}|^2$, and ${\cal R}(\eta)=3(C''/C^2-C'^2/C^3)$; 
for the case $C(\eta)=a^2(\eta)$ the curvature reads  ${\cal R}=6(a''/a^3)$.
The  normalization condition for $\chi_{\vec{k}}$ is given by \cite{mw}
\begin{align}
W(\chi_{\vec{k}}^{*},\chi_{\vec{k}})\ =\ \chi_{\vec{k}} \chi_{\vec{k}}^{*\prime} - \chi_{\vec{k}}^{*} \chi_{\vec{k}}^{\prime}\  =\  2i \ .
\label{Wronskian}
\end{align}
The  functions $\chi_{\vec{k}} $  satisfy the  initial conditions  at a particular moment of time $\eta_i$, 
being the preferred mode functions which determine the
(initial) vacuum state, or the lowest energy state.

In this paper, we will consider a {\em non-minimal} coupling scalar field, i.e., $\xi=1/6$. 
In this case, the frequency (\ref{Frequency}) reduces to
\begin{align}
\Omega_{k}^2(\eta)\ =\ k^2+m^2 a^2(\eta)\ .
\label{FrequencyB}
\end{align}
In order to find the evolution of quantum fields in our  cosmological background in all times, 
we need to consider both cosmological regions (i.e., $0<t<t_0$ and $t_0<t<t_b$) as described in section \ref{lqc}.

It is difficult to solve the Klein-Gordon equation (\ref{Master-Eq1})
for the general forms (\ref{scalefactor1}) and (\ref{scalefactor2-b}) of 
the scale factor, so that we will consider  
some simplifications. 
To do that, we first divide the entire evolution of the
universe into two phases: one which characterizes the
``primordial  bounce" phase ($t\rightarrow0$), for which it is possible to solve the
Klein-Gordon equation, so that, the solution naturally contains
the structure of the vacuum state of quantum fields (say, the `in' region).
The other which characterizes
the ``late time bounce" phase ($t\rightarrow t_b$). 
Then, we will  investigate the possibility of the  particle creation at the late time bounce (i.e., the `out' region).  
\begin{enumerate}[label=(\roman*)]
\item In the `primordial phase',  close to the initial bounce where $t \rightarrow 0$, 
one may expand the general expression of the scale factor  (\ref{scalefactor1})  as
\begin{align}
a(t) \  =\  a_i \left(1 + s^2t^2 \right) + {\cal O}(t^4)\ ,
\label{scalefactor1b}
\end{align}
where $s^2:=\kappa\rho_{\rm crit}/4$.
In terms of the conformal time $\eta$, by using  the relation $a^{-1}dt=d\eta$, we can rewrite the scale factor (\ref{scalefactor1b}) as
\begin{align}
a(\eta)\ \approx \ a_i\sec^2\left(sa_i \eta\right)\ .
\label{Sfactor-f1}
\end{align}
This indicates that as $t\rightarrow0$, then $\eta\rightarrow\eta_i=0$ at the initial bounce; $a(\eta_i=0)=a_i$.
Then, for small $\eta$, close to the initial bounce, we can expand Eq.~(\ref{Sfactor-f1}) and take the  first order terms:
\begin{align}
a(\eta)\ \approx \ a_i + s^2a_i^3 \eta^2\ ,
\label{prim-sc-a}
\end{align}
from which we have (to the first order terms)
\begin{align}
\Omega_{k}^2(\eta) \  \approx\  \omega_i^2 + 2m^2a_i^4s^2 \eta^2 \ .
\label{FrequencyB2-1}
\end{align}
where $\omega_i^2:=k^2+m^2a_{i}^2$. 
For later convenience, let us define new variables $x:=\sqrt{2\sqrt{2}ms}a_i\eta$ and $\tilde{\omega}_i^2=\omega_i^2/2\sqrt{2}msa_i^2$; in terms of these variables, 
we rewrite the Kelin-Gordon equation (\ref{Master-Eq1}) for the frequency (\ref{FrequencyB2-1}) as
\begin{align}
\frac{\partial^2\chi_{\vec{k}}}{\partial x^2} +\left(\tilde{\omega}_i^2+\frac{x^2}{4}\right)\chi_{\vec{k}}(x)=0\ .
\label{Master-Eq1-varNew}
\end{align}
This equation presents a parabolic cylinder equation whose solution can be written as a combination of hypergeometric functions 
$_1F_1(p, q, ix^2/2)$  \cite{Handbook}: 
\begin{align}
 \quad \quad \chi_{\vec{k}}(x)  \  & =\    e^{-i\frac{x^2}{4}}\left[c_1~_1F_1\left(p_1, q_1, ix^2/2\right) \right. \notag \\ 
&\ \quad  \quad   \quad   \left. + c_2x~_1F_1\left(p_2, q_2, ix^2/2\right)\right],
\label{field-alpha=2}
\end{align}
where $c_{1,2}$ are constants, $p_1=(i\tilde{\omega}_i^2/2+1/4)$, ~$q_1=1/2$; and $p_2=(i\tilde{\omega}_i^2/2+3/4)$,~  $q_2=3/2$.
Let us rewrite the solution (\ref{field-alpha=2}) in the form 
\begin{align}
\quad \chi_{\vec{k}}(x)\ =\ c_1 \chi_{\vec{k}}^{(1)}(x) + c_2 \chi_{\vec{k}}^{(2)} (x)\ .
\label{field-decomposition}
\end{align}
Then, by expanding the hypergeometric functions $_1F_1(p, q, ix^2/2)$ as series  \cite{Handbook}, we obtain 
\begin{align}
& \quad \quad  \chi_{\vec{k}}^{(1)}  = 1 - \tilde{\omega}_i^2\frac{x^2}{2!} - \left(\frac{1}{2} - \tilde{\omega}_i^4\right)\frac{x^4}{4!} + \cdot\cdot\cdot \\
& \quad \quad  \chi_{\vec{k}}^{(2)}  = x - \tilde{\omega}_i^2\frac{x^3}{3!} - \left(\frac{3}{2} - \tilde{\omega}_i^4\right)\frac{x^5}{5!} + \cdot\cdot\cdot
\end{align}
Considering  only to the order $x^3$ for  odd  series (and $x^2$ for even series), 
Eq.~(\ref{field-decomposition}) reduces to
\begin{align}
\quad \chi_{\vec{k}}(x)\ \approx\ c_1\cos(\tilde{\omega}_ix) + \frac{c_2}{\tilde{\omega}_i} \sin(\tilde{\omega}_ix)\ .
\label{field-decomposition2}
\end{align}
By choosing $c_1=1/\sqrt{2\omega_i}$ and $c_2 = i\tilde{\omega}_i/\sqrt{2\omega_i}$, 
we can rewrite (\ref{field-decomposition2}), in terms of the conformal time $\eta$, as
a typical normalized quantum vacuum state:
\begin{align}
\quad \chi_{\vec{k}}(\eta)\ = \   \frac{1}{\sqrt{2\omega_i}}e^{i\omega_i\eta}\  ,
\label{field-decomposition3}
\end{align}
which is in fact the initial vacuum state at $\eta\rightarrow\eta_i=0$.
Moreover, with the same choices of $c_{1,2}$ in Eq.~(\ref{field-alpha=2}) 
we have the solution for the quantum modes at any time.
Notice that, the modes (\ref{field-decomposition3}) are the negative-frequency solution with respect to $\eta$, 
therefore, all modes with equal $\omega_i$ are complex solutions of  
Eq.~(\ref{Master-Eq1}) such that, for a chosen  mode function $v_{k}(\eta)$ of that equation, 
the general solution $\chi_{\vec{k}}(\eta)$ can be expressed as  \cite{Birrell}
\begin{align}
\quad \quad \chi_{\vec{k}}(\eta)\ =\ \frac{1}{\sqrt{2}}\left[A_{\vec{k}}^{-}v_{k}(\eta)+A_{-\vec{k}}^{+}v_{k}^\ast(\eta)\right]\ ,
\label{mode-decompose1}
\end{align}
where $A_{\vec{k}}^{\pm}$ are constants of integration (being the so-called   
``annihilation" and ``creation" operators with respect to the mode $\chi_{\vec{k}}$ in quantum theory), and the mode function $v_k(\eta)$
is given by
\begin{align}
v_{k}(\eta)\ = \   \frac{1}{\sqrt{\omega_i}}e^{-i\omega_i\eta} \ ,
\label{field-decomposition3B}
\end{align}
which is  normalized due to $W[v_k, v_k^\ast]=2i$ given in Eq.~(\ref{Wronskian}).
Imposing  the reality condition   $\chi_{\vec{k}}^\ast(\eta)=\chi_{-\vec{k}}(\eta)$  into  
Eq.~(\ref{mode-decompose1}) we get $A_{\vec{k}}^{+}=(A_{\vec{k}}^{-})^\ast$; 
then by using  Fourier series  we can express the expansion of $\chi(\eta, \vec{x})$ as
\begin{align}
\chi(\eta, \vec{x}) \  =\  \frac{1}{(2\pi)^{\frac{3}{2}}} \int d^3k~ \chi_{\vec{k}}(\eta, \vec{x})\ ,
\label{mode-decompose2}
\end{align}
where $\chi_{\vec{k}}(\eta, \vec{x})$ is given by
\begin{align}
\quad \quad \chi_{\vec{k}}(\eta, \vec{x}) =
A_{\vec{k}}^{-}\frac{v_{k}(\eta)}{\sqrt{2}}e^{i \vec{k} \cdot \vec{x}} + A_{\vec{k}}^{+}\frac{v_{k}^\ast(\eta)}{\sqrt{2}}e^{-i \vec{k} \cdot \vec{x}}.
\label{mode-decompose2b}
\end{align}
In  the summation above  we have changed  the  variables  as $\vec{k}\rightarrow -\vec{k}$.

\item  At  the `late time phase'  as  $t\rightarrow t_b$,   the scale factor (\ref{scalefactor2-b}) can be approximated  as
\begin{align}
\quad \quad \bar{a}(\tau) \  = \  a_{b}\left(1 - \tilde{s}^2\tau^2\right) + {\cal O}\big(\tau^4\big)\ ,
\label{scalefactor2-approx-2}
\end{align}
where $\tau:=t-t_b$ and $\tilde{s}^2:=\frac{\kappa}{12}\rho_{\rm crit}|\beta|$.  Replacing  $\bar{a}(t)$  in  $\bar{a}d\tau=d\eta$  
from Eq.~(\ref{scalefactor2-approx-2}), by integration we obtain  the scale factor  in terms of the  proper time:
\begin{align}
\bar{a}(\eta) \ \approx \ a_b - \tilde{s}^2a_b^3(\eta-\eta_b)^2 \ ,
\label{scalefactor2-b-Final-1}
\end{align}
where $\eta_b$ is the proper time at which the universe hits a bounce at late time. 
The Eq.~(\ref{scalefactor2-b-Final-1})  looks   like the primordial scale factor (\ref{prim-sc-a}), 
so that, with a similar analysis we  
obtain the scalar field as  solution to  Klein-Gordon equation.
In particular, for  (\ref{scalefactor2-b-Final-1})  the frequency  (when considering only  first order terms) reads 
\begin{align}
\quad \Omega_{k}^2(\eta) \  \approx\  \omega_b^2 -  2m^2a_b^4\tilde{s}^2 (\eta-\eta_b)^2  \ .
\label{FrequencyB2-1-late}
\end{align}
where $\omega_b^2:=k^2+m^2a_{b}^2$. 
By introducing  $y:=\sqrt{2\sqrt{2}m\tilde{s}}a_b(\eta-\eta_b)$,  
the Klein-Gordon equation  for the frequency (\ref{FrequencyB2-1-late}),  in terms of  new variable  reads 
\begin{align}
\frac{\partial^2\chi_{\vec{k}}}{\partial y^2} -  \left(\frac{y^2}{4} - \tilde{\omega}_b^2\right)\chi_{\vec{k}}(y)=0\ ,
\label{KG-bounce2}
\end{align}
where $\tilde{\omega}_b^2:=\omega_b^2/2\sqrt{2}m\tilde{s}a_b^2$.
The differential equation (\ref{KG-bounce2}), as before,  is  a  parabolic cylinder equation  whose  solutions can be expressed as \cite{Handbook}
\begin{align}
& \quad \quad  \chi_{\vec{k}}^{(1)}  = 1 - \tilde{\omega}_b^2\frac{y^2}{2!} + \left(\frac{1}{2} + \tilde{\omega}_b^4\right)\frac{y^4}{4!} + \cdot\cdot\cdot \label{sol-1b} \\
& \quad \quad  \chi_{\vec{k}}^{(2)}  = y - \tilde{\omega}_b^2\frac{y^3}{3!} + \left(\frac{3}{2} + \tilde{\omega}_b^4\right)\frac{y^5}{5!} + \cdot\cdot\cdot \label{sol-2b}
\end{align}
Taking  only  the orders  $x^3$ for  odd  series and $x^2$ for even series, we 
express  the  total solution  to (\ref{KG-bounce2})   as a combination  of (\ref{sol-1b}) and (\ref{sol-2b}):
\begin{align}
\quad \chi_{\vec{k}}(y)\ \approx\ \tilde{c}_1\cos(\tilde{\omega}_by) + \frac{\tilde{c}_2}{\tilde{\omega}_b} \sin(\tilde{\omega}_by)\ ,
\label{field-decomposition2-latetime}
\end{align}
for arbitrary constants  of integration $\tilde{c}_1$ and $\tilde{c}_2$. For the choices of $\tilde{c}_1=1/\sqrt{2\omega_b}$ and $\tilde{c}_2=i\tilde{\omega}_b/\sqrt{2\omega_b}$, Eq.~(\ref{field-decomposition2-latetime})
reduces to $\chi_{\vec{k}}(\eta)\sim e^{i\omega_b(\eta-\eta_b)}/\sqrt{2\omega_b}$, indicating a (negative-frequency solution) vacuum state at  late time (i.e., the `out' region).

For each mode, the general solution to the Klein-Gordon equation (\ref{Master-Eq1}),  
in terms of the mode function $u_{k}(\eta)$ of the  `out' vacuum,  can be expanded, by using the new Bogolyubov coefficients  as  \cite{Birrell}
\begin{align}
\quad \quad \chi_{\vec{k}}(\eta)\ =\ \frac{1}{\sqrt{2}}\left[B_{\vec{k}}^{-}u_{k}(\eta) + B_{-\vec{k}}^{+}u_{k}^\ast(\eta)\right]\ ,
\label{mode-decompose1-sec}
\end{align}
in which a new complete orthonormal set of (positive-frequency) modes $u_{k}$ was considered as
\begin{align}
u_{k}(\eta)\ = \   \frac{1}{\sqrt{\omega_b}}e^{-i\omega_b(\eta-\eta_b)} \ .
\label{field-decomposition3B-ph2}
\end{align}
Notice that, similar to the `in' region in the primordial phase,  $u_k(\eta)$ herein this  phase, defines  the final vacuum state in the `out' region as $\eta\rightarrow\eta_b$. 
The $B_{\vec{k}}^{\pm}$ are constants of integration;  similarly they are  the  annihilation  and  creation operators in quantum theory, but,  with respect to the new vacuum state $u_{k}(\eta)$.
Therefore,  similar to (\ref{mode-decompose2b}) we can express the solution $\chi_{\vec{k}}(\eta, \vec{x})$ as
\begin{align}
\quad \quad \chi_{\vec{k}}(\eta, \vec{x}) =
B_{\vec{k}}^{-}\frac{u_{k}(\eta)}{\sqrt{2}}e^{i \vec{k} \cdot \vec{x}} + B_{\vec{k}}^{+}\frac{u_{k}^\ast(\eta)}{\sqrt{2}}e^{-i \vec{k} \cdot \vec{x}},
\label{mode-decompose2b-2}
\end{align}
where we again  applied  $\vec{k}\rightarrow -\vec{k}$ in the second term.
\end{enumerate}

\begin{figure}[t]
\includegraphics[height=2.1in]{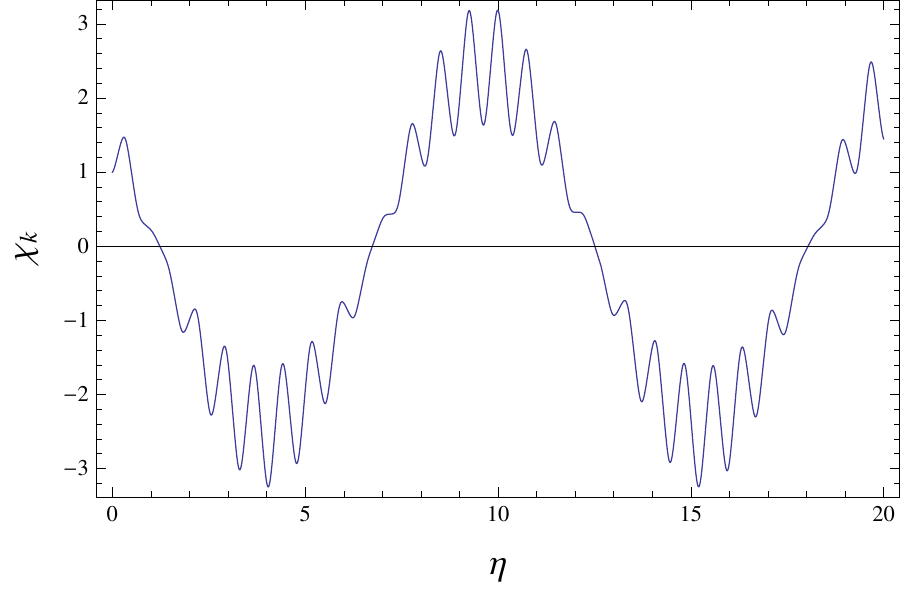} 
\caption{Time evolution   of the quantum field mode $\chi_k(\eta)$, given by the Klein-Gordon equation (\ref{Master-Eq1}), evolving between the initial and final bounces. 
We considered the following rescaled  choices of parameters in our numerical analysis: $\beta=3(1+w_d)=-6.6$, $\rho_0/\rho_{\rm crit} = 0.005$, $a_i=0.56$, $\dot{a}(0)=\dot{a}_i=1$, $k=0.5$ 
and $m=0.001$ in Planck units.
}
\label{fig-field} 
\end{figure}
Figure  \ref{fig-field}  shows typical numerical solution to the Klein-Gordon equation  (\ref{Master-Eq1})  for the
field in the herein cyclic cosmological model, for sub-planckian values of  the mass $m$
and wavenumbers $k$. It presents an oscillatory behaviour for  scalar
field evolving in the universe between the initial and final bounces. When
those parameters become trans-planckians, the oscillations give places to asymptotically
divergent  behaviour, as could be expected.

Frequencies (\ref{FrequencyB2-1}) and   (\ref{FrequencyB2-1-late}) are the same for a massless field, thus, the solutions (\ref{mode-decompose2b}) and (\ref{mode-decompose2b-2})
can be written in the form of the plane waves in two phases. Since the solutions must be continuous, there is no final effect in the 
final bouncing phase for particle production. 
However, in the massive case situation is totally different.
In order to discuss this  more precisely, 
 we will calculate,  in the following section, the Bogolyubov coefficients at the final 
 bounce in the presence of   non-unique  vacuum solutions in the `in' and `out' regions.

\subsection{The Bogolyubov transformations}

In quantum theory, the mode operator $\hat{\chi}_{\vec{k}}(\eta, \vec{x})$
can be expressed through the creation and annihilation operators \cite{Birrell}:
in the primordial phase, where $\eta\rightarrow 0$ (denoted by `in' region), 
constants  $A_{\vec{k}}^{\pm}$ are denoted to be the annihilation and creation operators.
Thus, replacing this  in Eq.~(\ref{mode-decompose2b}) we obtain the mode operators:
\begin{align}
\hat{\chi}_{\vec{k}}^{\rm (in)}(\eta, \vec{x})=
\hat{A}_{\vec{k}}^{-}\frac{v_{k}(\eta)}{\sqrt{2}}e^{i \vec{k} \cdot \vec{x}} + \hat{A}_{\vec{k}}^{+}\frac{v_{k}^\ast(\eta)}{\sqrt{2}}e^{-i \vec{k} \cdot \vec{x}}.
\label{mode-decompose2b-1a}
\end{align}
Similarly, for the late time bounce phase, where $\eta\rightarrow \eta_b$ (denoted  by  `out' region), from Eq.~(\ref{mode-decompose2b-2}) we obtain the mode operators as
\begin{align}
\hat{\chi}_{\vec{k}}^{\rm (out)}(\eta, \vec{x})=
\hat{B}_{\vec{k}}^{-}\frac{u_{k}(\eta)}{\sqrt{2}}e^{i \vec{k} \cdot \vec{x}} + \hat{B}_{\vec{k}}^{+}\frac{u_{k}^\ast(\eta)}{\sqrt{2}}e^{-i \vec{k} \cdot \vec{x}}.
\label{mode-decompose2b-2a}
\end{align}

The expansions (\ref{mode-decompose2b-1a}) and (\ref{mode-decompose2b-2a}) express the same field $\hat{\chi}_{\vec{k}}(\eta, \vec{x})$ through two different
sets of functions, so the $k$-th Fourier components of these expansions must agree
\begin{align}
\hat{\chi}_{\vec{k}}^{\rm (in)}(\eta_c, \vec{x})\ =\ \hat{\chi}_{\vec{k}}^{\rm (out)}(\eta_c, \vec{x})\ ,
\label{match1}
\end{align}
at any transition time $\eta=\eta_c$. (We can consider $\eta_c=\eta_0$ to be the transition time from matter dominant region to the dark energy dominant region.) 
Furthermore, continuity condition at $\eta=\eta_c$, from one era to another implies that the (conformal) time derivative of 
mode functions must satisfy
\begin{align}
\frac{\partial \hat{\chi}_{\vec{k}}^{\rm (in)}(\eta_c, \vec{x})}{\partial\eta} \ =\  \frac{\partial \hat{\chi}_{\vec{k}}^{\rm (out)}(\eta_c, \vec{x})}{\partial\eta}\ .
\label{match2}
\end{align}
From the matching conditions (\ref{match1}) and (\ref{match2}) we obtain
\begin{align}
\hat{B}_{\vec{k}}^{-} \ & = \  \alpha_k \hat{A}_{\vec{k}}^{-} + \beta_k^\ast \hat{A}_{-\vec{k}}^{+}\ ,
\end{align}
and
\begin{align}
\hat{B}_{\vec{k}}^{+} \ & = \     \alpha_k^\ast \hat{A}_{\vec{k}}^{+}  +  \beta_k \hat{A}_{-\vec{k}}^{-}  \ .
\end{align}
These relations  present  the Bogolyubov
transformations in which the old annihilation and creation operators $\hat{A}_{\vec{k}}^{\pm}$ are expressed through the new operators $\hat{B}_{\vec{k}}^{\pm}$, 
with the $\eta$-independent (complex) Bogolyubov coefficients: 
\begin{align}
\alpha_{k} \ & = \ \frac{\sqrt{\omega_b}}{2\sqrt{\omega_i}}\left(1 + \frac{\omega_i}{\omega_b}\right)e^{-i(\omega_i-\omega_b)\eta_c -  i\omega_b\eta_b}\ ,  \\
\beta_{k} \ & = \ \frac{\sqrt{\omega_b}}{2\sqrt{\omega_i}}\left(1 - \frac{\omega_i}{\omega_b}\right)e^{-i(\omega_i+\omega_b)\eta_c  + i\omega_b\eta_b}\  .
\end{align}
Using these coefficients we can express the primordial basis $v_k(\eta)$ in terms of the late time one, $u_k(\eta)$, as
\begin{align}
v_k(\eta)\ =\ \alpha_k u_k(\eta) + \beta_k u^\ast_k(\eta)\ .
\end{align}
Since $v_k(\eta)$ and $u_k(\eta)$ are normalized, it follows that
\begin{align}
\alpha_k\alpha_k^\ast - \beta_k\beta_k^\ast = 1\ .
\end{align}
The coefficient $\beta_k$ is associated with the created particles.
More precisely, in the Heisenberg picture, the initial vacuum state $|0_{\rm in}\rangle$ in the `in' region is the state of the system for all time.
The physical number operator which counts particles in the `out' region reads  $\hat{N}_k=\hat{B}_{\vec{k}}^+\hat{B}_{\vec{k}}^-$  \cite{Birrell}.
Then,  the mean number of $u_k$-mode particles in the state $|0_{\rm in}\rangle$ is given by  
\begin{align}
{\cal N}_k \ & =\  \langle 0_{\rm in}| \hat{N}_k | 0_{\rm in} \rangle \ \notag \\
& =\ \beta_k\beta^\ast_k \ =\ \frac{\omega_b}{4\omega_i}\left(1 - \frac{\omega_i}{\omega_b}\right)^2 ,
\label{particle-number1}
\end{align}
which is to say that the vacuum state of  the $v_k$ modes contains ${\cal N}_k$ particles in the $u_k$ mode.
For a massless scalar field, $\omega_i=\omega_b=|\vec{k}|$, thus, $\alpha_k=1$, $\beta_k=0$ and the mean density ${\cal N}_k$ is zero;
therefore, no particle is  produced  in the case of a massless 
scalar field in the  `out' region.

\subsection{The energy density  of created particles}

The relation (\ref{particle-number1}) is  the average number of late time particles per spatial volume and per wave number $k$. 
The  energy density of the $\ell$ particles in the vacuum
state $|0_{\rm in}\rangle$ for each mode $v_k$ reads $\rho_k=(\frac{1}{2}+\ell)\omega_i$, 
where $w_i/2$ is the zero-point energy.
Since the chosen  quantum state  corresponds to the `in' vacuum  $|0_{\rm in}\rangle$, then
in the `out' region ($\eta\rightarrow\eta_b$) the combined mean density of total particles for each mode $k$ 
(after subtracting the zero-point energy) is $\rho_k={\cal N}_k\omega_i$ \cite{Birrell:1980}.
Then, for all  modes (per unit volume), we obtain the density of late time particles as
\begin{align}
\rho_P\ &=\  \int_0^\infty \frac{d^3k}{(2\pi)^3} |\beta_k|^2\omega_{i}  \notag \\
& =\   \int_0^\infty  \frac{dk}{8\pi^2}~ k^2 \omega_b\left(1 - \frac{\omega_i}{\omega_b}\right)^2\ .
\label{particle-density}
\end{align}
At small scales close to the initial bounce (`in' region)  we can  approximate  $ma_i\ll 1$,  thus,  $\omega_i\approx k$.  Then, Eq.~(\ref{particle-density}) reduces to
\begin{align}
\rho_P\ &\approx \  \frac{1}{8\pi^2} \int  dk~ k^2 \omega_b \left(1 - \frac{k}{\omega_b}\right)^2 \notag \\
& = \  \frac{1}{8\pi^2}\left[\omega_bk\left(\frac{k^2}{2}-\frac{m^2a_b^2}{4}\right) \right. \notag \\
& \quad  \quad  \quad  \quad   \left.  - \frac{k^4}{2} + \frac{m^2a_b^2}{4}\ln(2k+2\omega_b)\right].
\label{particle-density-massless}
\end{align}
This integral has obviously,  a logarithmic divergence when it tends to the ultraviolet limit,  $k\rightarrow \infty$ \cite{Julio:2008,Julio:2011}, 
so that  it is necessary to regularize it.

Using the $n$-wave method developed in Ref.~\cite{Pauli-Villars}, which consists in subtracting terms obtained by expanding $\rho_k$ in powers of $k^{-2}$: 
\begin{align}
\rho_k^{\rm ren} \ =\  \rho_k  -  E_k^{(0)} -  E_k^{(1)} - \frac{1}{2} E_k^{(2)}
\end{align}
in which we have defined
\begin{align}
E^{(p)}_k \ =\ \lim_{n\rightarrow\infty} \frac{\partial^p \rho_k^{(n)}}{\partial(n^{-2})^p}\ ,
\end{align}
where $\rho_k^{(n)}:= \rho_k(nk, nm)/n$,
and $n$ is the parameter that characterizes the order of the divergence. 
Therefore, $E_k^{(0)}$, $E_k^{(1)}$ and $E_k^{(2)}$ eliminate, respectively, the logarithmic,  quadratic and the quartic divergences.
Following the regularisation procedure corresponding to a
full renormalisation of the coupling constants \cite{Grib:1994,Grib:1998}, we have that
\begin{align}
\rho_k  \ =\  \omega_b - 2k + \frac{k^2}{\omega_b}
\end{align}
which follows that $\rho_k^{(n)}=\rho_k$; then, we find the  renormalised energy as \cite{Julio:2011}
\begin{align}
\rho_k^{\rm ren} \ =\ \rho_k - E^0_k \ = \  0 \ ,
\end{align}
which is zero. Vanishing energy density of created particles indicate that 
quantum field effects associated with the cosmological dynamics at late time do not change the nature of
quantum gravity bounce in the far future.

\section{Conclusion and Discussion}
\label{conclusion}

In this work, we considered a flat FLRW universe whose matter content is dominated by a dust fluid at early time and a  
phantom dark energy component at late time (cf. see Section \ref{lqc}). 
We   employed  loop quantum cosmology  to govern  the geometry of  space-time at Planck era.
In particular, quantum gravity  effects modified the dynamics of the universe in the primordial  and late time phases,
thus, resolved the  big bang singularity  in the past and  the big rip singularity in the  future,  and replaced them by quantum bounces.
This indicated that the resulting quantum bounce and recollapse in our herein model contribute a cyclic scenario for the universe.
In classical cosmology,  in general,  different matter components   provide  different types  of singular fate for the universe at late times.  
Subsequently, loop quantum gravity  resolves  those singularities so  that the resulting quantum bounce and recollapse give rise to different cyclic scenarios  for the universe.
Our main concern in this paper was to construct a cyclic scenario in  the FLRW universe, in order to investigate the phenomena of
cosmological particle production by  quantum gravity effects.  Nevertheless,  for  further  studies  of   possible cyclic  models    driven by loop quantum cosmology, 
we suggest  readers  to refer to  the recent paper Ref.~\cite{Singh:2011} and references therein.

We  studied  quantum theory of  a {\em massive}, non-minimally coupled scalar field  on the resulting cyclic   cosmological  background.
Our  aim was to investigate  particle production by  semiclassical gravitational effects,  in an expanding  phase   towards  late time.
In general,  in a curved space-time  background, it is not possible  to  define a unique  vacuum state \cite{Parker:1974}. 
However,  in the  asymptotically static   `in' or  `out'  regions followed by a period of expansion,
the notion of  asymptotic  vacuum  states  come   closest to a Minkowski-like vacuum  \cite{Birrell}. 
In our herein model, the scale factor of the universe starts expanding from an initial quantum bounce (for some times $\eta_i\sim0$,
on  $a\sim a_i$,  where $a_i\sim (M/m_{\rm Pl})^{1/3}\ell_{\rm Pl}$ and $M$ is the total  mass of the universe) 
 towards a late time bounce (for some times $\eta\sim\eta_b$, where $a\sim a_b=a_0(\rho_{\rm crit}/\rho_0)^{1/|\beta|}$).
Therefore,  the scale factor of the universe admits two asymptotic static  forms at the initial and final states%
\footnote{Similar assumption was previously employed for a universe hitting a  {\em sudden} singularity in the far future \cite{Julio:2008,Julio:2011}. 
Therein, asymptotic   static  (out)  region was considered to be at the singular phase ($t\sim t_s$), 
where    the scale factor and its  first time derivative approach constant values, while  the second time derivative of the scale factor diverges as $t\rightarrow t_s$.
Moreover,  the transition from one regime to the other, at $\eta_c$, was assumed to be at the singular phase, $a(\eta_c)=a_s$ and $\dot{a}(\eta_c)=\dot{a}_s=const.$}.
We have used this static approximation,  as  `in' or  `out'  regions,  in the computation  of  created particles (cf.  Section \ref{QFT}). 
However, it must be stressed that,  this does not lead to a  trivial situation, since this approximation is different
from the usual Minkowski asymptotic approximation \cite{Birrell} and from the sudden
singularity approximation \cite{Julio:2008,Julio:2011}: in the present case, 
 the transition from a matter dominated region  to a  dark energy dominated region, 
gave rise to the   Planck regimes (in which quantum gravity effects are  inevitable)  
at  two different asymptotic phases, 
which consequently, leaded to   distinct  `in' and `out'  vacuum states.
Moreover, in the asymptotic limits 
(i.e., $\eta\rightarrow0$ and $\eta\rightarrow\eta_b$), the first time derivative of
the scale factor is zero, but the second time derivative is finite. 
These different
contexts are reflected in the matching conditions (via transition time $\eta_c$) connecting the initial and
final states. Nevertheless,  it does not affects on the final particle production, 
since the particle number (cf. see   Eq.~(\ref{particle-number1})) depends only on the characteristic of the initial and final vacuum.

For a  scalar field,  undergoing cosmological evolution between the initial and final static phases,
we solved  the Klein-Gordon equation  in  the (past) matter dominated region ($\eta<\eta_0$),  and  
the (future)  dark energy dominated region ($\eta>\eta_0$);  
the conformal time  $\eta_0$ denotes the {\em present}, being  the transition time (i.e., $\eta_c=\eta_0$) from matter dominated regime to  dark energy dominated regime.
We  obtained  mode  functions  of the fields  close to   the initial and final  bounces whose forms are similar to that  of  a flat space mode function.
For the herein {\em massive} scalar field, different cosmic scales  of  the initial and final phases 
leaded to  different frequencies for the modes  of the quantum fields, and consequently,  
provided  non-unique  vacuum states   in  the `in' and `out'  regions;
this  non-uniqueness of the vacuum   was accompanied by  particle production phenomena  in the `out' region.
For a {\em massless} scalar field, however, mode functions of  the  `in' and `out' regions are the same  as  $e^{-ik\eta}$, possessing a unique vacuum of the  Minkowski space; 
hence,  there is no final effects for  the particle production at late time, for  massless fields.  

The  energy density of quantum scalar field, $\rho_\varphi$,   in the herein  model, 
was assumed to be negligible initially in comparison to the quantum  background density (which is of the order $\rho\sim\rho_{\rm crit}=0.41\rho_{\rm Pl}$) near the initial bounce. 
Moreover,  there were  no particles production  initially  since $A_{\vec{k}}^+A_{\vec{k}}^-|0_{\rm in}\rangle=0$.  
Our concern, thus,  was  to investigate the back-reaction of quantum  effects carried  by  the total particles created  in the `out' region. 
We  thus computed  the total density of the  quantum particles, $\rho_P$,  created  at  late time,
on approach to the future bounce where the energy density, curvature and the scale factor of the background geometry  reaches their upper  bounds.
Our analysis showed that the density of created particles in the `out'  region   remains  negligible  in comparison to   the quantum background density (i.e,  $\rho_P\ll\rho_{\rm crit}$).
This   indicates    that the back-reaction effects of the quantum particle production do not influence  the  evolution of the universe  until late time.
In addition,   the future bounce, induced   by quantum gravity effects,  was   unscathed by the back-reaction from the quantum field. 

One important issue for the model analysed here  is the possibility of observational traces of the quantum phase and of the oscillatory behaviour of the universe.
This would require, for example, to compute the evolution of perturbations in the model. However, we must face an important limitation:  
We have no classical covariant formulation of the equations of motion governing the evolution of the universe. Strictly speaking, this analysis would require to return to
the full quantum model in performing the perturbative analysis, what implies in considerable technical difficulties  (c.f. Refs.~\cite{Ashtekar:2013,Andrea:2013}).
This drawback is less severe in the case of gravitational waves  (see  Ref.~\cite{Bojowald:GW}), since there is no coupling of the spin 2 modes with the scalar and vectorial modes (which
are directly connected to the matter content), at least at linear level. If the gravitational waves, represented by the propagation of the traceless, transverse field $\delta g_{ij} = h_{ij}$,
obeys the same equation as in the usual general relativity case, that is,
\begin{eqnarray}
\mu''  + \Big(k^2 - \frac{a''}{a}\Big)\mu\  =\  0, 
\label{gw}
\end{eqnarray}
where $h_{ij} = a\epsilon_{ij}\mu$, with $\epsilon_{ij}$ being the polarisation tensor and $a(\eta)$ the scale factor, 
we would have the same equation as for a massless scalar field propagating
in the geometry determined by the background equations \cite{Grishchuk:1993}. 
We would have in this case, qualitatively, the same behaviour as displayed in figure \ref{fig-field}  for
the massive field, but with an accumulative divergence for transplackian values of $k$. 
However, it is not sure that equation (\ref{gw}) remains the same in the present case. 
Moreover, the transplanckian regime would require a full quantum
analysis, including for the background. Due to these important difficulties to surmount, 
we postpone a perturbative analysis for future researches.

In general, a   cyclic universe seems to be in  conflict with  the  second law of thermodynamics.
More precisely,  the  entropy of the universe  increases,  through natural thermodynamic processes,  in every cycle of expansion and contraction  \cite{Tolman:1931,Tolman:1934}.
So that, at the beginning of a new cycle, there is higher entropy density (than the previous cycle)  because of the entropy    added  from earlier cycles.
It turns out that the duration of a cycle is sensitive to the entropy density, thus  by increasing the entropy, the duration of the cycle increases as well. 
On the other hand, the cycles become  shorter and shorter the further back we go
until, after a finite time, they shrink to zero duration.  
New hope for a consistent cyclic cosmology,   to  get rid of this puzzling situation, was provided
by introducing a  cyclic  model  based  on (phantom) dark energy \cite{Baum-Frampton:2007, Baum-Frampton:2008}, and also in 
an ekpyrotic  scenario  derived from  the  brane-world  cosmological model
\cite{Steinhardt-Turok}: 
in both scenarios, there are accelerating phases leading to  a suppression of entropy.
In the context of  LQC,  there are   several proposals  in order to deal with the issue of  entropy problem  
in   bouncing universe  \cite{Bojowald:2007,Bojowald:2008,Corichi:2008}. 
In addition,  presence of the quantum fields in such  bouncing  scenario
may  also give rise to the  significant effects on  entropy  generation  from  particle creation; 
 there are  inquiries on finding a viable measure 
of entropy for particle production  processes in literature  (see for example Refs. \cite{Hu:1986,Hu:1987}).
In the present model, the oscillatory behaviour indicated  in the quantum modes (see figure  \ref{fig-field}) may be a hint that there is no major problem
with the cumulative effects as the cycle  goes on.
This oscillatory  behaviour breaks down as the Planck regime is approached.
But, the approximation made in the present work may not be valid in  the deep Planck regime.
The necessary renormalization procedure changes also the high frequency behaviour.
Nevertheless, more investigation is needed in order to
understand  the effects of particle creation on generation of the entropy near  the quantum bounce in the full quantum theory.

\section*{ACKNOWLEDGMENTS}

YT thanks  CNPq (Brazil) for financial support.
JCF also thanks CNPq (Brazil) and FAPES (Brazil) for  partial financial support.
This work was also supported by the project CERN/FP/123609/2011.


\end{document}